# Textured growth and electrical characterization of Zinc Sulfide on back-end-of-the-line (BEOL) compatible substrates

Claire Wu[1], Mythili Surendran[2, 3], Anika Tabassum Priyoti[4], Gokul Anilkumar[1], Cheng-Hsien Wu[5], Chun-Chen Wang[5], Cheng-Chen Kuo[5], Harish Kumarasubramanian[1], Kenta Lin[4], Amari Butler[1], Rehan Kapadia[4], Xinyu Bao[5], Jayakanth Ravichandran[1, 3, 4]

1. Mork Family Department of Chemical Engineering and Materials Science, University of Southern California, Los Angeles, California 90089, USA
2. Lawrence Berkeley National Laboratory, 1 Cyclotron Road, Berkeley, California 94720, USA
3. Core Center for Excellence in Nano Imaging, University of Southern California, Los Angeles, California 90089, USA
4. Ming Hsieh Department of Electrical and Computer Engineering, University of Southern California, Los Angeles, California 90089, USA
5. Corporate Research, Taiwan Semiconductor Manufacturing Company, Ltd., Hsinchu, Taiwan R.O.C.

Scaling of transistors has enabled continuous improvement in the performance of logic devices, especially with contributions from materials engineering. However, there is a need to surpass the horizontal limitations in chip manufacturing and incorporate the vertical or third dimension. To enable monolithic three-dimensional (M3D) integration of high-performance logic, one needs to solve the fundamental challenge of low temperature (< 450°C) synthesis of high mobility *n*-type and *p*-type semiconductor thin films that can be utilized for fabrication of back-end-of-line (BEOL) compatible transistors[1]. Metal oxides, especially indium oxides alloyed with gallium and tungsten[2,3], are promising *n*-type semiconductor channel materials; however there is a lack of *p*-type channel materials that can meet the stringent synthesis conditions of BEOL manufacturing. Zinc sulfide (ZnS), a transparent wide band-gap semiconductor, has shown room temperature *p*-type conductivity when doped with copper and nitrogen, and crystallizes below 400°C.[4] Here, we report growth of crystalline thin films of ZnS by pulsed laser deposition (PLD) on a variety of amorphous and polycrystalline surfaces including silicon nitride ($SiN_x$), thermal silicon dioxide ($SiO_2$), yttrium oxide ($Y_2O_3$), hafnium dioxide ($HfO_2$), sapphire ($Al_2O_3$), platinum (Pt), and titanium nitride (TiN). X-ray diffraction scans show out-of-the-plane texturing of ZnS on all surfaces. In-plane crystalline quality is investigated using grazing incidence wide-angle X-ray scattering measurements. Surface and interface quality is measured using X-ray reflectivity and atomic force microscopy measurements. Electrical characterization of the ZnS films is done by J-V measurements of ZnS on platinum and metal-oxide-semiconductor capacitor (MOSCAP) measurements of ZnS on $SiO_2$ on heavily doped silicon. The J-V measurements indicate low leakage current on the order of $10^{-5}$ A/cm$^{-2}$ with electric field of 0.40 MV cm$^{-1}$ and the MOSCAP characteristics show bilayer capacitor behavior, which points to ZnS being highly intrinsic with very low unintentional, electrically active point defects. Further work on doping ZnS with copper or other *p*-type candidate dopants are needed to demonstrate ZnS as a dopable wide band gap semiconductor for channels compatible with BEOL manufacturing. This work showcases the capability of novel thin film growth technique of a wide band-gap sulfide semiconductor in BEOL compatible conditions with potential for technological applications in transistor manufacturing.

# I. INTRODUCTION

As the scaling of conventional silicon-based field effect transistors is approaching the dimensional limit in 2D, future hardware for computing and information technologies turn to heterogeneous 3D integration of high-performance integrated circuits (ICs), where memory, logic, and photonic functions shall be incorporated into the back-end-of-the-line (BEOL) manufacturing[5]. BEOL manufacturing has a stringent synthesis temperature requirement of <450°C and hence, beyond silicon technologies are needed for the semiconducting channels[6]. Today several candidates (*e.g.*, organic polymers, semiconducting oxides, amorphous or polycrystalline silicon, III-nitrides)[7–11] are actively investigated, but currently none of the candidates fulfill all the necessary requirements, such as the ability to form high mobility *n*- and *p*-type channels, low temperature processing, low voltage operations, substrate adaptability and reliability. Although numerous review articles have discussed advancements in *p*-type oxide semiconductors[12], there is a notable gap in the literature concerning *p*-type sulfide semiconductors, where predominantly layered dichalcogenides are the widely studied candidates. Zinc sulfide has been widely reported as a candidate material for photodetectors, display applications and as a passivation layer for $Cd_{1-x}Zn_xTe$[13–15]; however, only recently has it been reported as a viable *p*-type semiconductor[4]. Synthesis of sulfide thin films presents challenges such as managing cation-chalcogen vapor pressure mismatch, as well as slow diffusion kinetics and use of highly corrosive and reactive chalcogen precursors that can degrade the film-substrate interface[16,17]. To address the challenges of thin film growth of vapor-pressure mismatched materials and to mitigate rough surfaces and interfaces, we developed a novel hybrid PLD (*h*PLD) method[18], and we use this technique to synthesize thin films of ZnS in this report. This technique uses an organosulfur precursor as a sulfur source in conjunction with PLD, without the flow of a background gas, to produce stoichiometric films. Additionally, the sulfur precursor decomposes at a lower temperature and with higher efficiency than other sources such as elemental sulfur and Ar-$H_2S$[19–22]. Hybrid PLD has been shown as an effective technique for textured growth of sulfide thin films, and electrical properties are highly dependent on the texture of the film[23–26]. Thus, it is crucial to use a controlled method to achieve highly textured films and their corresponding electrical properties such as capacitance, carrier concentration, and carrier mobility. In this article, we report the growth of textured zinc sulfide thin films, by hybrid pulsed laser deposition (PLD) on several BEOL compatible amorphous and polycrystalline surfaces as well as electrical characterization of the films by I-V and MOSCAP measurements.

# II. EXPERIMENTAL

## A. Thin film deposition

The thin films were grown by *h*PLD method using a 248 nm KrF excimer laser in a vacuum chamber (Demcon TSST) specifically designed for the growth of chalcogenides. The chamber was evacuated to a base pressure of ~ $10^{-8}$ mbar. The targets for ablation were densified using either sintering or cold isostatic pressing. High purity stoichiometric ZnS (Alfa Aesar, 99.99%) powder was pressed into pellets with a diameter of ¾ inch and densified to >90% density by room temperature cold isostatic pressing. $Y_2O_3$ (Alfa Aesar, 99.99%) and $HfO_2$ (Materion 99.9%) powder were pressed into ¾ inch pellets and densified to >90% density by sintering at 1550°C for 48 hours. These dense pellets were used as the targets and were pre-ablated before growth. Prior to deposition, silicon substrates (University Wafer) were etched in hydrofluoric acid buffered oxide etched (HF BOE) solution (1:50 HF BOE:$H_2O$) for 2 minutes to remove native oxide layer and rinsed with de-ionized water 5-6 times prior to deposition. The polycrystalline films of $Y_2O_3$, $HfO_2$, and TiN were deposited *via* PLD at 400°C onto HF BOE etched Si substrates and Pt was sputtered onto HF BOE etched Si (004) at room temperature. $Al_2O_3$ was deposited *via* atomic layer deposition (ALD) at 100°C and plasma enhanced chemical vapor deposition (PECVD) was used to deposit $SiO_2$ and $SiN_x$. The substrate was heated up to 400°C in vacuum prior to deposition of $Y_2O_3$, $HfO_2$, and TiN. TiN was grown in nominal vacuum at a background pressure of ~$10^{-6}$ torr and $Y_2O_3$ and $HfO_2$ were grown in the presence of oxygen gas at a partial pressure of $10^{-4}$ torr and $10^{-2}$ torr, respectively. For ZnS, an organosulfur compound, tert-butyl disulfide (TBDS) (99.999%, Sigma–Aldrich) was used as the sulfur precursor for the PLD growth. Since TBDS is a liquid with a relatively high vapor pressure at room temperature, it was introduced by thermal evaporation in a stainless-steel bubbler that was connected through a gas inlet system to the growth chamber. The bubbler was heated to a temperature of 130-140°C using external heating tapes. The TBDS precursor delivery was controlled using a linear leak valve to achieve total chamber pressure of ~$10^{-3}$ Torr. No carrier gas was used. The fluence was fixed at 1.0-1.2 J $cm^{-2}$ and the target-substrate distance used was 75 mm. The films were cooled post growth at a rate of 10 °C $min^{-1}$ at an Ar-$H_2S$ (4.97 molar ppm $H_2S$) partial pressure of 100 mTorr.

## B. Structural and Surface Characterization

The powder X-ray diffraction (XRD) was carried out on a Bruker D8 Advance diffractometer using a soller slit with Cu $K_\alpha$ ($\lambda$ = 1.5418 Å) radiation at room temperature.

XRR measurements were performed on the same diffractometer in a parallel beam geometry using a Göbel (parabolic) mirror setup. GIWAXS experiments were conducted in vacuum conditions and at room temperature using a Xenocs Xeuss 3.0 laboratory instrument with an incident wavelength of 1.5406 Å. An incident angle of 0.5° was used for GIWAXS. X-ray scattering data were recorded on a Pilatus 300k area detector at a sample-detector distance of 72 mm. AFM was performed on Bruker Multimode 8 atomic force microscope in peak force tapping mode with a ScanAsyst tip geometry to obtain the surface morphology and roughness.

### *C. Electrical Measurements*

Circular electrodes of diameters from 70 μm to 470 μm were deposited using standard photolithography and thermal evaporation of 10 nm chromium and 40 nm gold. I-V measurements were conducted using a semiconductor parameter analyzer (Agilent 4156C) and the MOSCAP measurements were conducted using a Keysight B1500A semiconductor parameter analyzer. A probe station was used in a dark cabinet to avoid photoconductive effects.

## III. RESULTS AND DISCUSSION

Zinc sulfide thin films were grown from a phase pure stoichiometric ZnS target on structurally and chemically diverse, amorphous and polycrystalline surfaces such as dielectric and metallic materials using the *h*PLD approach. Due to high propensity for loss of sulfur at high temperatures, presence of excess sulfur during growth is essential to maintain stoichiometric films. Additionally, attaining phase pure, textured films at low temperatures could be challenging as previous reports of ZnS films grown by PLD have been either polycrystalline and not textured in a preferential orientation and/or grown at higher synthesis temperatures[27–34]. There have been reports of low temperature growth of ZnS films by other techniques such as spray pyrolysis and RF magnetron sputtering, but the films are polycrystalline without any preferential texture as well. These lower temperatures for the fabrication of BEOL devices are critical to maintain the integrity of front-end-of-the-line devices and interconnects. Here, we discuss the structure of textured ZnS films that were successfully grown at BEOL compatible temperatures (450°C or below) using *h*PLD.

## A. X-ray diffraction characterizing texture of films

Structural information obtained from powder diffraction studies was used as the primary criterion for investigating out-of-plane texture. All the characterization was performed on as-grown films without any further annealing. In Fig. 1 (a), we show representative X-ray diffraction patterns of ZnS thin films with weak out-of-plane texture on polycrystalline $Y_2O_3$, $HfO_2$, Pt, and TiN films. The ZnS films show *c*-axis oriented texture as indicated by the 00*l* reflections in XRD scans in Fig. 1(a). As shown in Fig. 1(b), thin films of ZnS show similar weak out-of plane texture on amorphous surfaces of $Al_2O_3$, $SiO_2$, $SiN_x$ and directly on HF BOE silicon. Bragg reflections at 61.83°, 65.98°, 66.54°, 69.27°, and 69.47° correspond to the 400 reflection of the silicon substrate corresponding to the X-ray wavelengths of Cu such as Cu-K$\beta_1$, W-L$\alpha_1$, W-L$\alpha_2$, Cu-K$\alpha_1$, Cu-K$\alpha_2$, respectively. Despite being a forbidden reflection, due to the multiple diffraction depending on the azimuthal angle chosen, 002 Bragg reflection of Si appears at 34.6° in certain XRD patterns in Fig. 1 (a) and (b)[35]. The texturing of ZnS on a variety of surfaces and substrates indicates that film-substrate lattice matching is not required to texture ZnS thin films grown by *h*PLD.

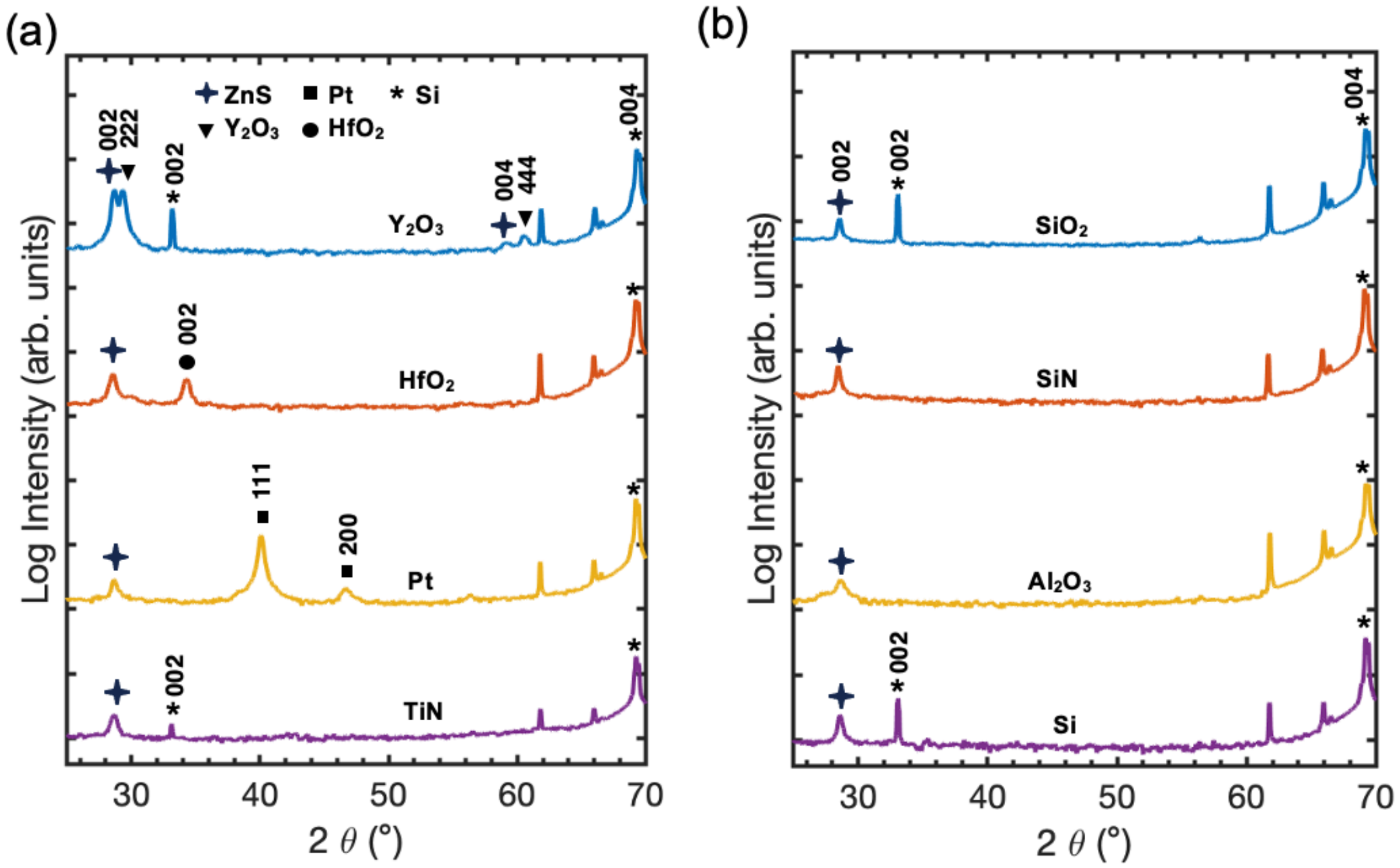

**FIG 1**: (a) Powder diffraction $\theta$-$2\theta$ XRD patterns of representative ZnS films on polycrystalline surfaces of TiN, Pt, $HfO_2$, and $Y_2O_3$ and (b) amorphous surfaces of $SiO_2$, $SiN_x$, $Al_2O_3$, and directly deposited on HF BOE etched silicon substrate.

## B. X-ray reflectivity and AFM

To characterize the surface quality and interfaces of the films, X-ray reflectivity (XRR) and atomic force microscope (AFM) measurements were performed. ZnS films of 25 nm thickness were grown on the polycrystalline surfaces of $Y_2O_3$, $HfO_2$ and TiN. The thicknesses of the ZnS films grown on $Al_2O_3$, $SiO_2$, $SiN_x$, and Pt that are shown here were 30 nm, 92 nm, 58 nm, and 200 nm, respectively. The XRR and AFM studies on the same films are shown in Fig. 2. The slow decay of the reflected X-ray intensity and the presence of Kiessig fringes indicate that both the polycrystalline templating layer and the ZnS films have smooth surfaces and interfaces. The X-ray intensity for the ZnS thin film on Pt decays more quickly than the other polycrystalline films due to the roughness of the sputtered platinum of ~1.5 nm. ZnS films grown on amorphous surfaces such as $SiO_2$ and $SiN_x$ have increased interface roughness in comparison to films grown on polycrystalline surfaces. This aligns with the deposition technique used for the amorphous surfaces since films produced from sputtering (Pt) and PECVD ($SiO_2$ and $SiN_x$) have higher values of roughness of 3-5 nm and porosity. However, AFM topography images in Fig. 2 (d) show that the roughness of ZnS film grown on $SiN_x$ surface is lower than the $SiN_x$ amorphous template (Fig. 2 (c)), indicating a smooth conformal film with sufficient surface diffusivity to smoothen the surface roughness of the template. $SiN_x$ has a root mean squared (RMS) roughness of 4.14 nm and the ZnS film grown on top has an RMS roughness of 1.71 nm. A similar trend was observed for ZnS film with RMS roughness 2.43 nm grown on amorphous $SiO_2$ with RMS roughness 1.34 nm (Fig. S2 (b)). The RMS roughness for ZnS films grown on $Al_2O_3$ and TiN are 0.81 nm and 1.71 nm, respectively (Fig. S2 (c), (d)). The consistent smoothness of ZnS films on a variety of surfaces indicates that the *h*PLD growth technique can be leveraged to enhance surface diffusivity at lower temperatures and on a variety of surfaces to grow textured ZnS. Next, we will discuss structural characterization to understand the in-plane nature of the ZnS films, and electrical properties to correlate the crystallinity and point defects in ZnS thin films.

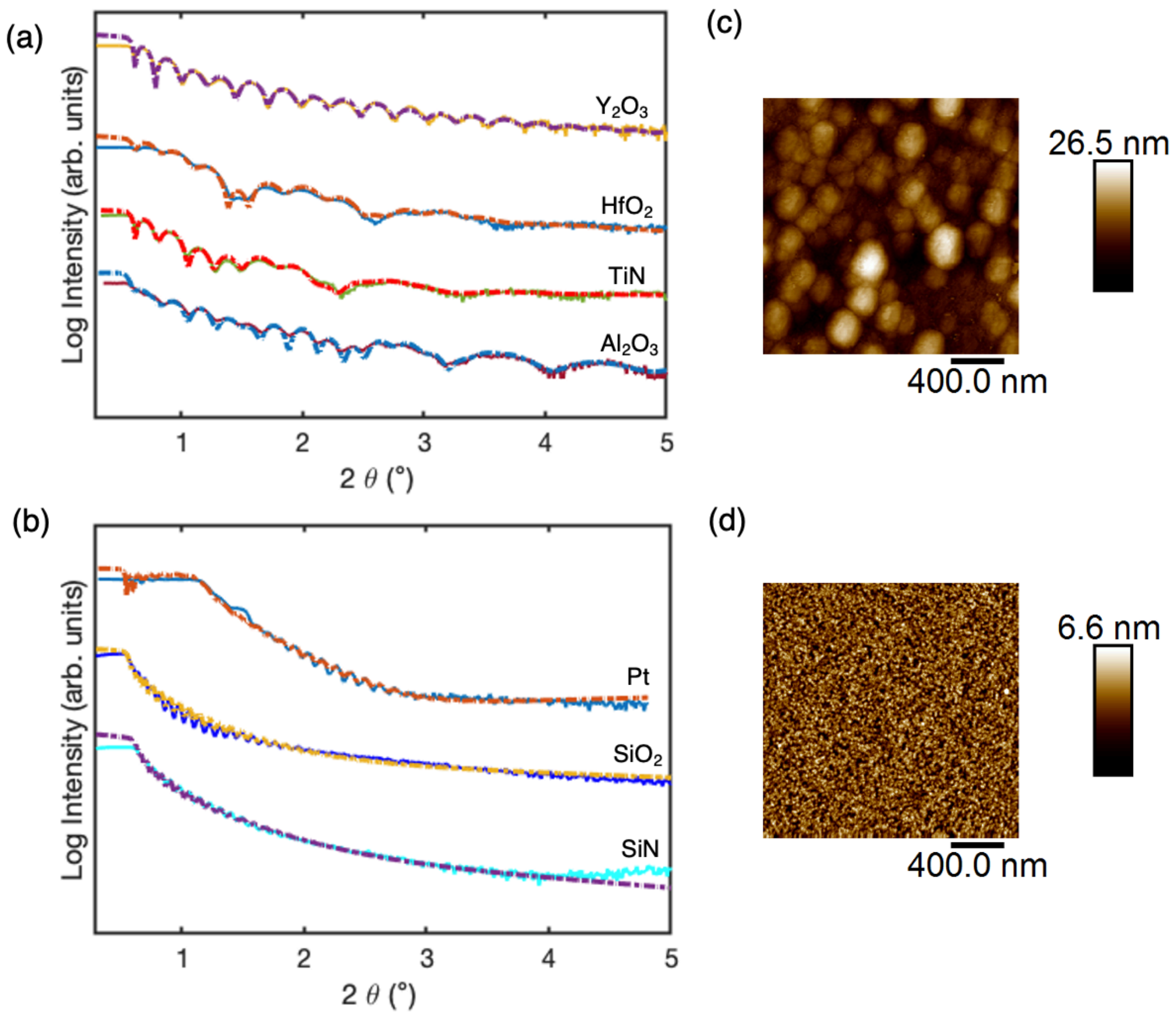

**FIG 2.** (a) XRR patterns showing 30 nm ZnS films grown on polycrystalline surfaces of $Y_2O_3$, $HfO_2$, TiN, and $Al_2O_3$ and (b) 120 nm ZnS film was grown on a polycrystalline surface of Pt, and 90 nm ZnS film grown on amorphous $SiO_2$, and $SiN_x$ on silicon substrates. The corresponding 2 µm x 2 µm AFM images of (c) PECVD amorphous $SiN_x$ film on silicon and (d) ZnS film on $SiN_x$ amorphous surface.

## *C. Polycrystalline in-plane orientation - GIWAXS*

HRXRD $2\boldsymbol{\theta}$-$\boldsymbol{\theta}$ scans were performed to investigate the texturing of ZnS films on amorphous and crystalline surfaces. However, the out-of-plane texturing of the film was too weak and the structural domain sizes were too small to elucidate diffraction peak intensities using laboratory scale diffractometers. Therefore, we used grazing incidence wide-angle X-ray scattering (GIWAXS), a conventional X-ray scattering technique, to determine in-plane film orientations of the ZnS thin films. The major advantage of the GIWAXS characterization is the 2D area detector that can acquire a simultaneous signal collection in both the in-plane and the out-of-plane directions much faster over a broader range of Q-values (Q refers to reciprocal space wave vector). Thus, it provides additional information about the texture of disordered, polycrystalline films. Fig. 3 (a) shows the GIWAXS pattern of 30 nm ZnS film on $Y_2O_3$/Si, a polycrystalline surface and Fig. 3 (b) shows the GIWAXS pattern of 35 nm ZnS on $SiN_x$/Si, an amorphous surface.

An inaccessible region or a "missing wedge" exists in the GIWAXS pattern in Fig. 3 due to the curvature of the Ewald sphere. There may be diffraction signals from highly ordered crystallites in the inaccessible hidden zone. In Fig. 3 (a), on $Y_2O_3$, we observe in-plane diffraction streaks of ZnS representing different grain orientations present. This indicates that ZnS has more appreciable preferential texturing on $Y_2O_3$ (111), a dielectric material, but not on TiN or Pt, which are other polycrystalline surfaces whose GIWAXS pattern is shown in Supplementary Fig. 3 and 6, respectively. To elucidate the nature of preferential texturing of ZnS on $Y_2O_3$, a GIWAXS measurement was taken on the $Y_2O_3$ template layer grown on etched silicon (Fig. S7). The pattern shows polycrystalline rings of $Y_2O_3$ representing multiple families of reflections with stronger texture in the 222 family of reflections at $Q = 2.05$ Å$^{-1}$, and the 125 family of reflections at $Q = 3.25$ Å$^{-1}$. This suggests that highly oriented grains of the templating layer can impact the in-plane grain orientations of ZnS. Several grain orientations were observed of ZnS, with 101 and 103 reflections as the most intense with partial rings and diffraction streaks indicating some degree of in-plane ordering. Other diffraction streaks corresponding to the 100, 102, 110, and 112 families of reflections are indexed as well. On $SiN_x$, we observe multiple polycrystalline rings at Q = 1.9 $Å^{-1}$ corresponding to lattice planes with a *d*-spacing value of 3.13 Å. This can be indexed as the ZnS 100 family of reflections. Similarly, the ring at Q = 2.76 $Å^{-1}$ has a *d*-spacing value of 2.28 Å and is indexed to the 102 family of reflections. The remaining polycrystalline rings are indexed as the 110 family of reflections at Q = 3.29 $Å^{-1}$, 103 family of reflections at Q = 3.56 $Å^{-1}$, and 112 family of reflections at Q = 3.85 $Å^{-1}$. These rings indicate no preferential in-plane grain orientations of the ZnS film on $SiN_x$. Given the amorphous and rough templating surface of $SiN_x$, the polycrystallinity of the ZnS film serves as contrast to the textured ZnS film on smooth, textured $Y_2O_3$. Additional GIWAXS measurements of the other BEOL compatible surfaces ($SiO_2$ & $Al_2O_3$) can be found in Fig. S4 & S5. The GIWAXS measurements of the ZnS films indicate randomly oriented in-plane orientations on all surfaces except for $Y_2O_3$, which has some in-plane ordering of ZnS grain orientations due to the texturing of $Y_2O_3$.

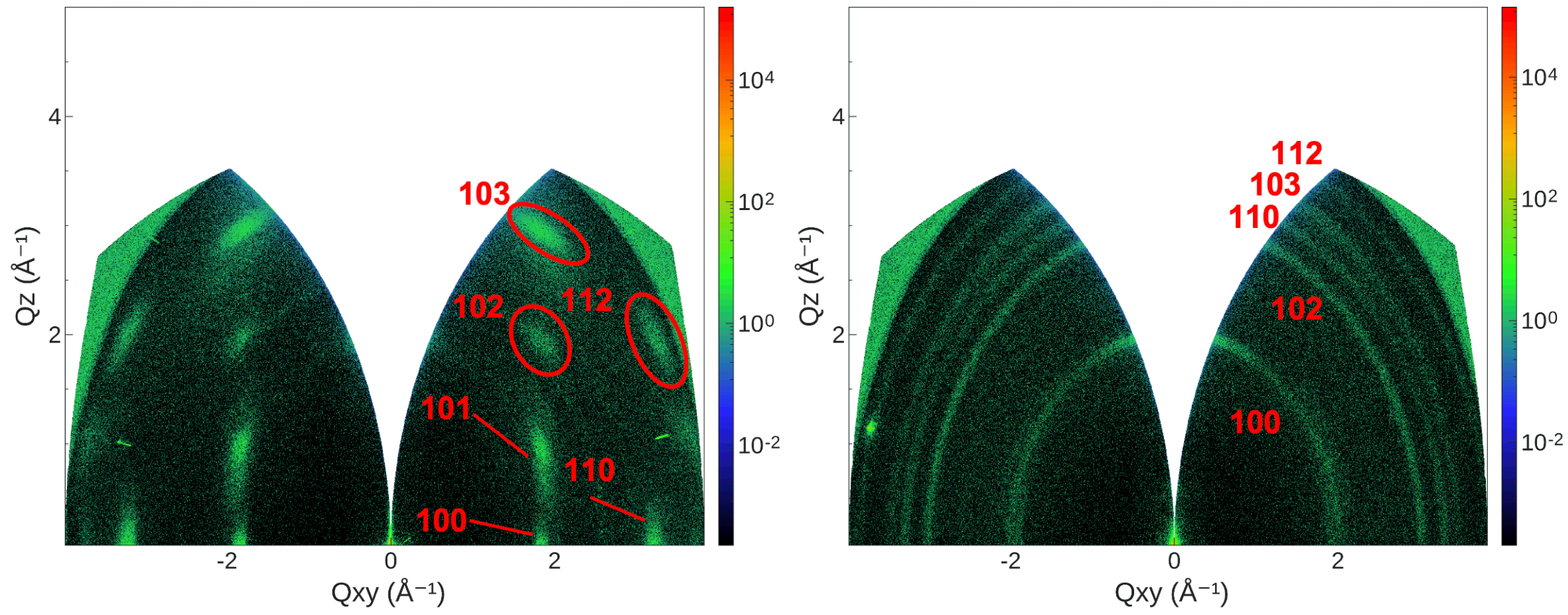

**FIG 3.** GIWAXS pattern for wurtzite ZnS film at an incident angle of 0.5° showing different diffraction streaks present on (a) $Y_2O_3$ polycrystalline film and ZnS film polycrystalline rings present on (b) amorphous $SiN_x$ film.

## *D. Electrical Characterization*

The electrical characterization was carried out by J-V and MOSCAP measurements at room temperature on ZnS thin films on thermally grown $SiO_2$. The J-V characteristics as shown in Fig. 4 (b) were used to reveal a leakage current per unit area on the order of $10^{-5}$ A $cm^{-2}$ at an electric field of 0.40 MV $cm^{-1}$. The capacitance-voltage (C-V) characteristics were tested at 2 kHz in a sweeping voltage range from -20 V to 20 V with a voltage step of 40 mV, a wait time of 0.1 s, and a hold time of 5 s, which aim at keeping MOS capacitors at steady state[36]. The flat C-V curve indicates that ZnS is a highly intrinsic material acting as bilayer capacitor with $SiO_2$ and does not demonstrate the typical capacitance trend expected from a doped semiconductor layer in a MOS structure. The relative dielectric constants of $SiO_2$ and ZnS were determined to be $\varepsilon_{SiO_2}$= 3.7 and $\varepsilon_{ZnS}$ = 5.6 based on the flat capacitance equation: $C_{ox} = \varepsilon_{ox}\varepsilon_0/d$, where d is the film thickness and by assuming ZnS and $SiO_2$ act as capacitor layers in series, $\frac{1}{C_{total}} = \frac{1}{C_{ZnS}} + \frac{1}{C_{SiO_2}}$. The dielectric constant of ZnS was determined from a ZnS film grown on platinum sputtered on heavily doped silicon in an anisotropic metal-insulator-metal structure. The top metal contacts deposited (Cr/Au) differ from the back contact of Pt/Si; thus, the work function between ZnS and the metal contacts are different. The J-V curve of ZnS/Pt/Si can be found in Fig. S8. Regarding the lower dielectric constants of ZnS (around 5.6) and $SiO_2$ (around 3.7) than the theoretical values of 8 and 3.9, respectively, which are determined by single layer dielectric capacitors, we make cautious inference that the issue is due to interface trapping effects[37,38]. The J-V measurements indicate low leakage current and the MOSCAP characteristics show bilayer capacitor behavior, which points to ZnS being highly intrinsic with very low unintentional, electrically active point defects. Further work on doping ZnS with copper or other *p*-type candidate dopants are needed to demonstrate ZnS as a dopable wide band gap semiconductor for channels compatible with BEOL manufacturing.

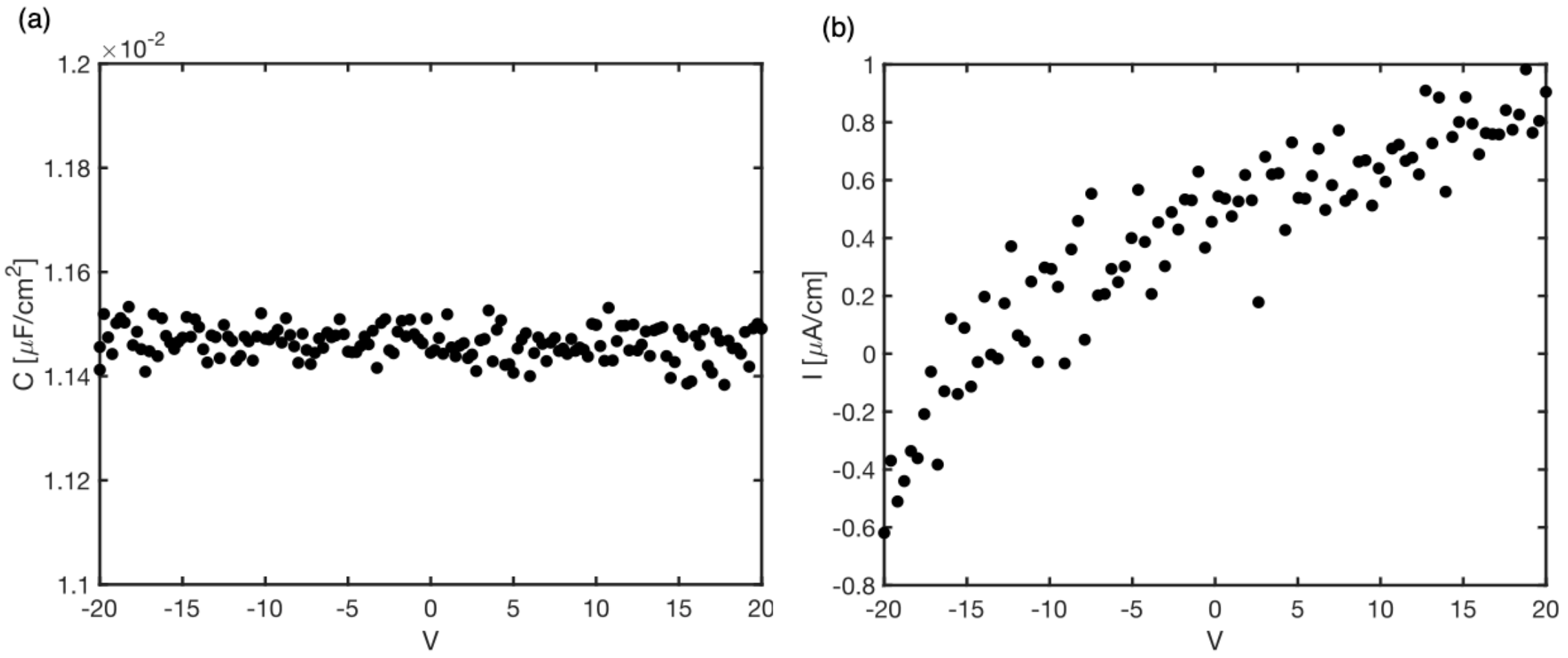

**FIG 4.** (a) Cross-plane C-V measurement and (b) in-plane I-V measurement of 100 nm ZnS film on amorphous $SiO_2$ on heavily doped silicon.

This characterization demonstrates the successful growth of ZnS thin films on chemically and structurally diverse surfaces using hybrid pulsed laser deposition (*h*PLD). The crystallinity of the films was characterized by XRD and GIWAXS to determine weak out-of-plane texture across various substrates. The surface that demonstrated the strongest in-plane grain orientations was polycrystalline $Y_2O_3$, while polycrystalline ZnS rings were observed on the remaining surfaces. AFM and XRR studies showed smooth ZnS films despite the intrinsic roughness of the amorphous surfaces. Electrical characterization indicated that ZnS has low leakage current and is a highly intrinsic wide band gap semiconductor that necessitates *p*-type dopants to induce MOSCAP behavior. These findings underscore the potential of hybrid PLD-grown ZnS thin films for back-end-of-line (BEOL) applications in transistor manufacturing.

# IV. SUMMARY AND CONCLUSIONS

Textured ZnS thin films have been demonstrated by *h*PLD on a variety of amorphous and polycrystalline surfaces in BEOL compatible conditions. We have demonstrated consistent growth of ZnS thin films in terms of structure (texture), roughness, and in-plane grain orientations, as measured by X-ray diffraction, X-ray reflectivity, atomic force microscopy, and grazing incidence wide-angle X-ray scattering. X-ray diffraction shows weak out-of-plane texture on all amorphous and polycrystalline surfaces. X-ray reflectivity and AFM show smooth films despite rough amorphous surfaces. GIWAXS studies show textured in-plane grain orientations of ZnS on the $Y_2O_3$ polycrystalline surface and no clear preferred in-plane orientations on all other surfaces. Electrical characterization shows low leakage current and flat capacitance curves, indicating the need for doping ZnS with *p*-type dopants to fabricate MOSCAP and MOSFET structures. This work emphasizes the ability of *h*PLD to synthesize ZnS as a *p*-type material in BEOL conditions for thin film 3D integration of transistor stack. Next, it is worthwhile to explore a variety of *p*-type dopants for ZnS and electrically characterize *h*PLD grown ZnS via capacitance and FET measurements to demonstrate the viability of ZnS as a *p*-type semiconductor for BEOL applications.

# SUPPLEMENTARY MATERIAL

## *A. Atomic Force Microscopy (AFM)*

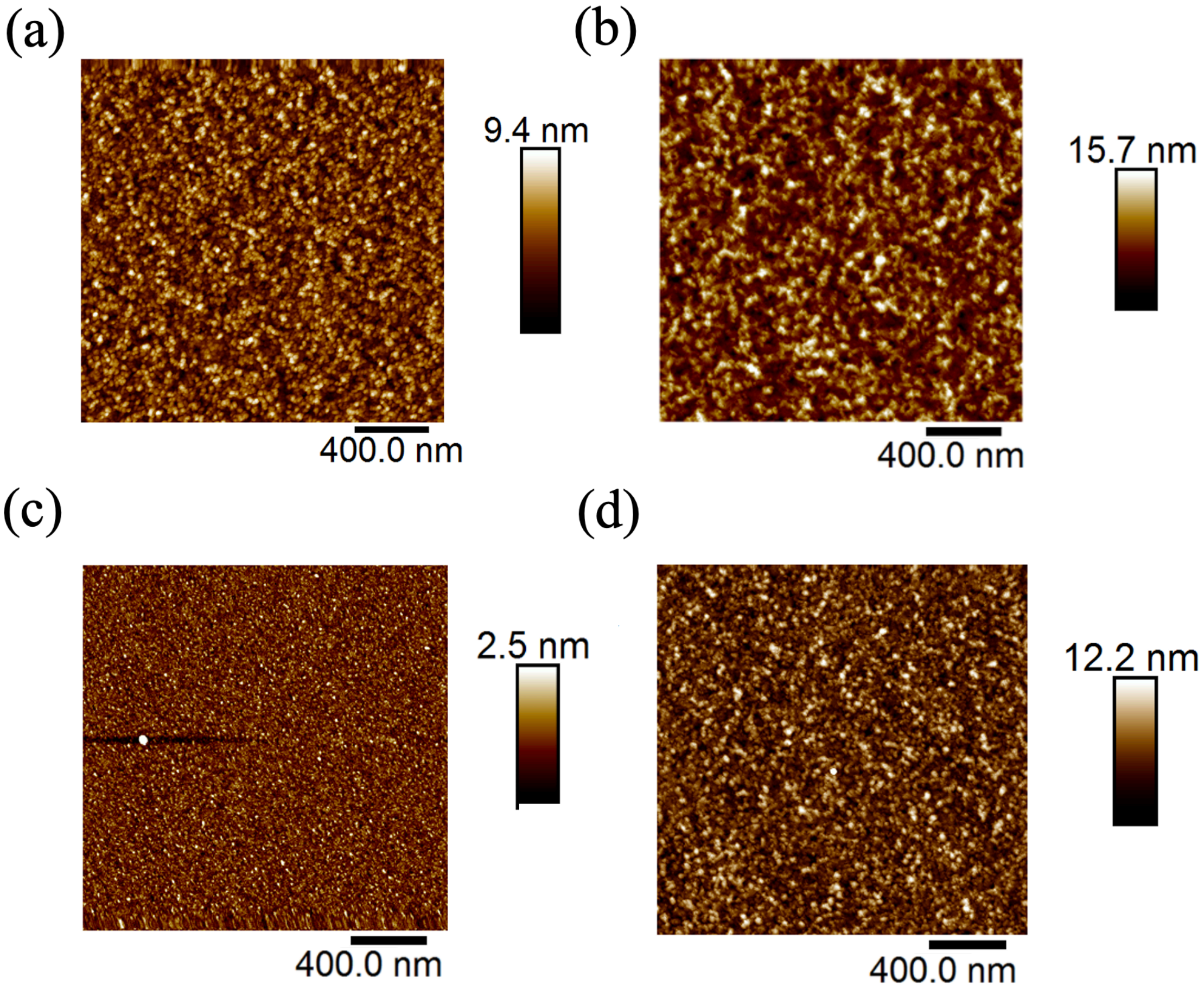

**FIG S1.** The corresponding AFM images of (a) 30 nm ZnS with RMS roughness of 1.53 nm on polycrystalline $HfO_2$/Si and (b) 90 nm of ZnS with RMS roughness of 2.43 nm on amorphous $SiO_2$/Si. (c) 20 nm of ZnS with RMS roughness of 0.81 nm on amorphous $Al_2O_3$/Si. (d) 25 nm of ZnS with RMS roughness of 1.71 nm on polycrystalline TiN/Si.

## *B. Grazing Incidence Wide Angle X-ray Scattering (GIWAXS)*

All GIWAXS measurements were taken at an incident angle of 0.5°.

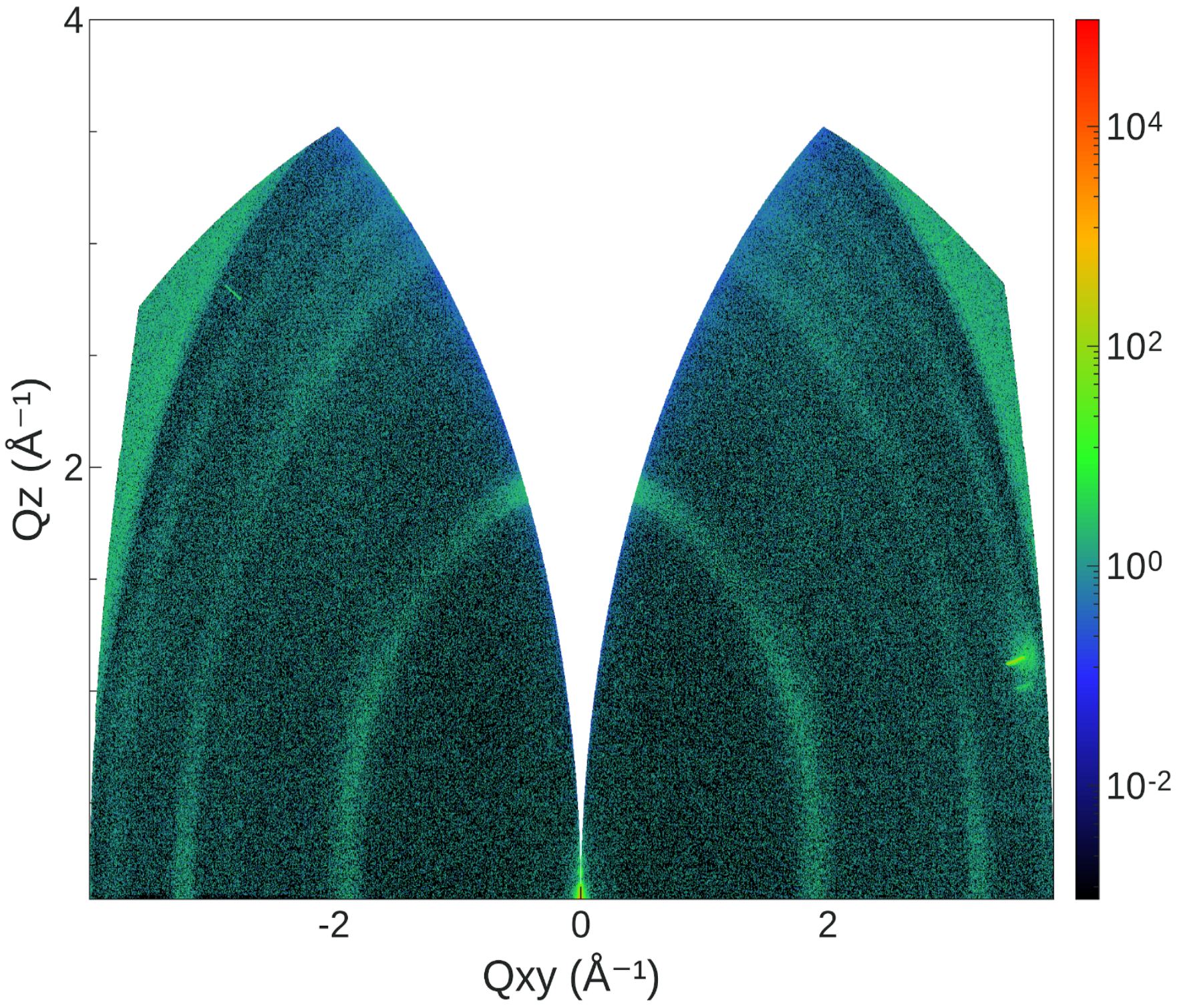

**FIG. S2.** GIWAXS pattern of ZnS film on TiN film deposited in vacuum by PLD on etched HF BOE Si. Polycrystalline rings at Q = 1.9 Å$^{-1}$, 3.13 Å$^{-1}$, 3.31 Å$^{-1}$ correspond to ZnS texturing with 100, 110, and 112 families of reflections, respectively.

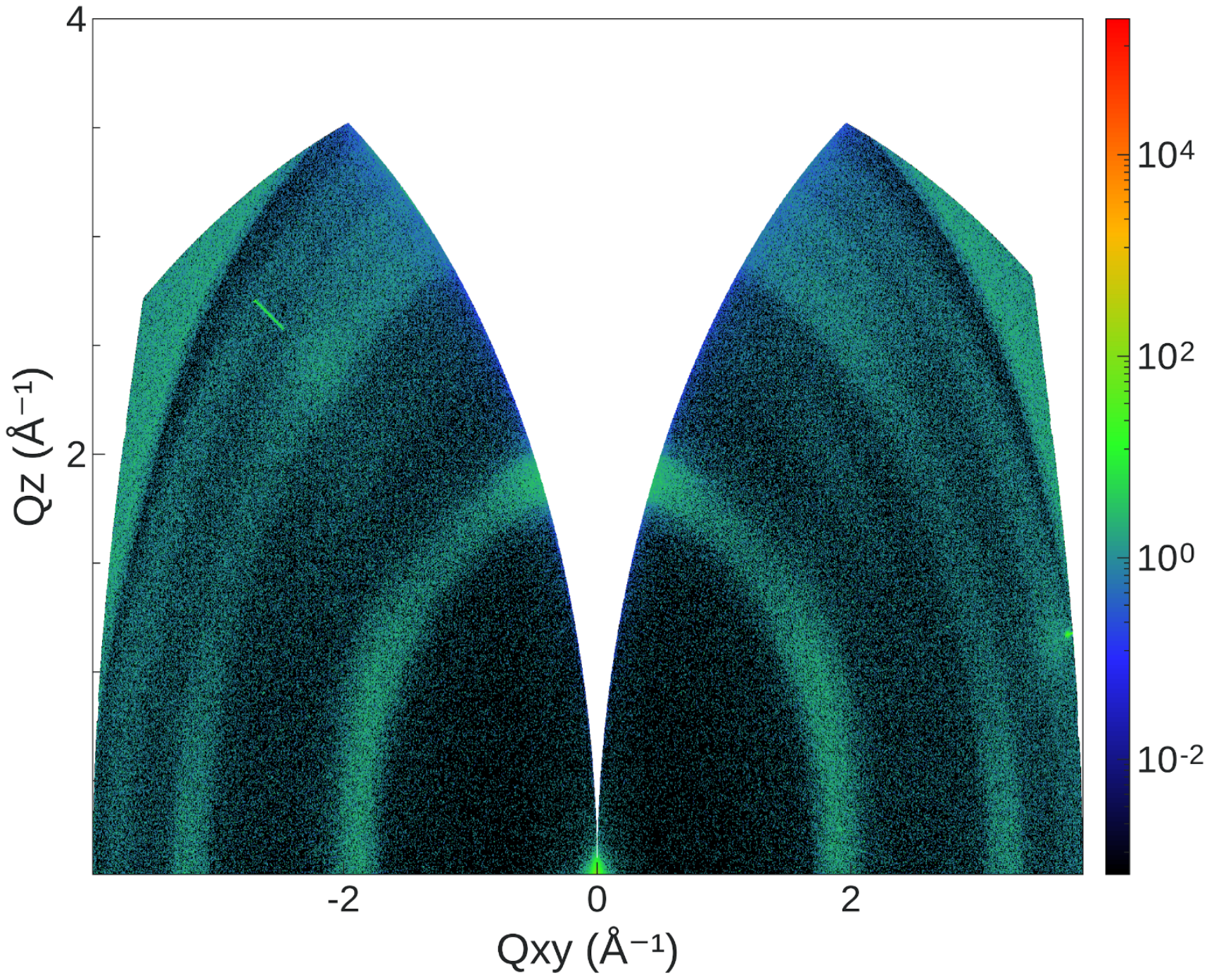

**FIG. S3**. GIWAXS pattern of 30nm ZnS film on $Al_2O_3$ film deposited by ALD on silicon. Polycrystalline rings at Q = 1.9 $Å^{-1}$, 3.13 $Å^{-1}$, 3.31 $Å^{-1}$ correspond to ZnS texturing in 100, 110, and 112 families of reflections, respectively.

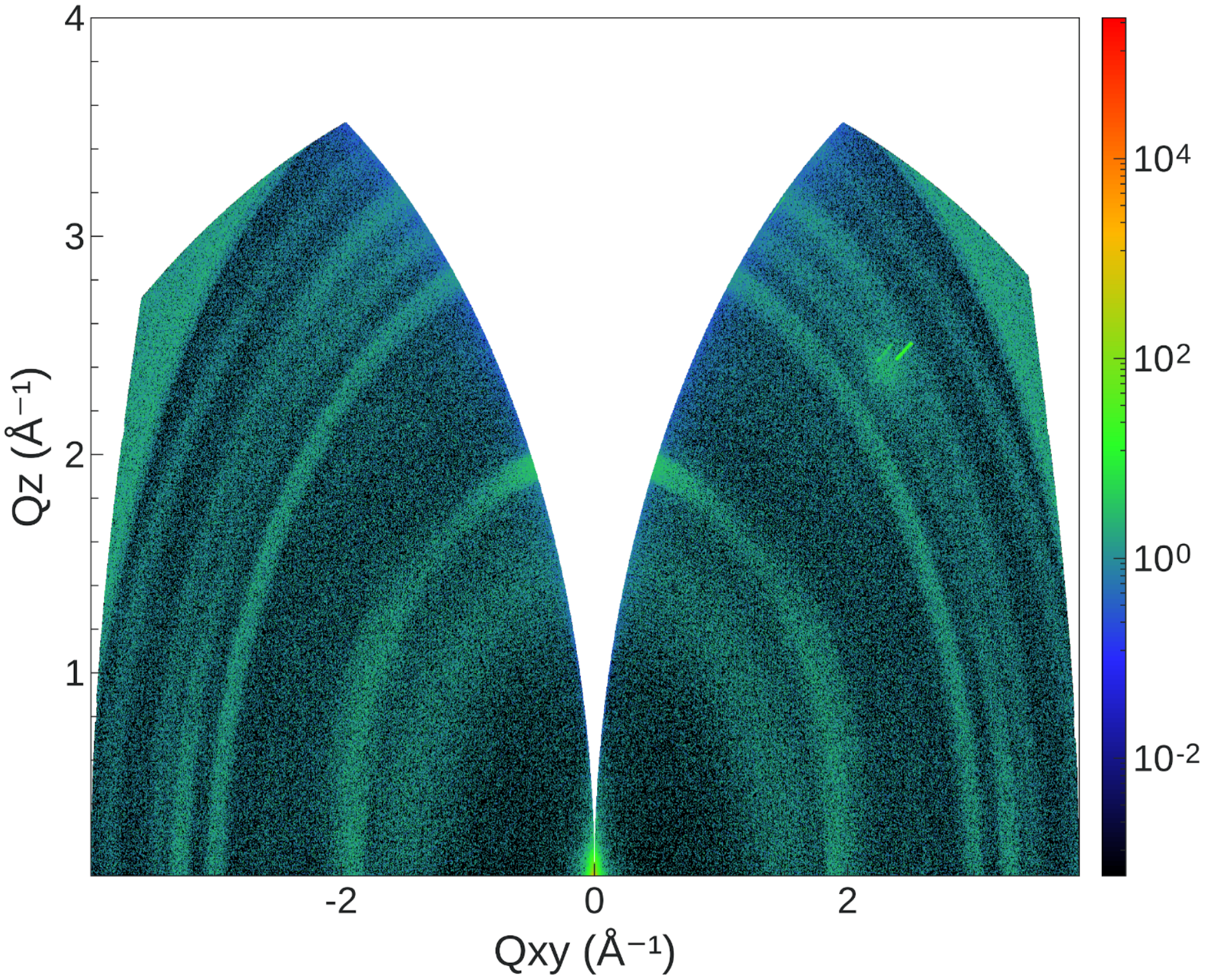

**FIG. S4.** GIWAXS pattern of 90 nm of ZnS on amorphous PECVD $SiO_2$ on silicon. Polycrystalline rings at Q = 1.9 $Å^{-1}$, 2.76 $Å^{-1}$, 3.29 $Å^{-1}$, 3.56 $Å^{-1}$, 3.85 $Å^{-1}$ correspond to ZnS texturing in 100, 102, 110, 103, and 112 families of reflections, respectively.

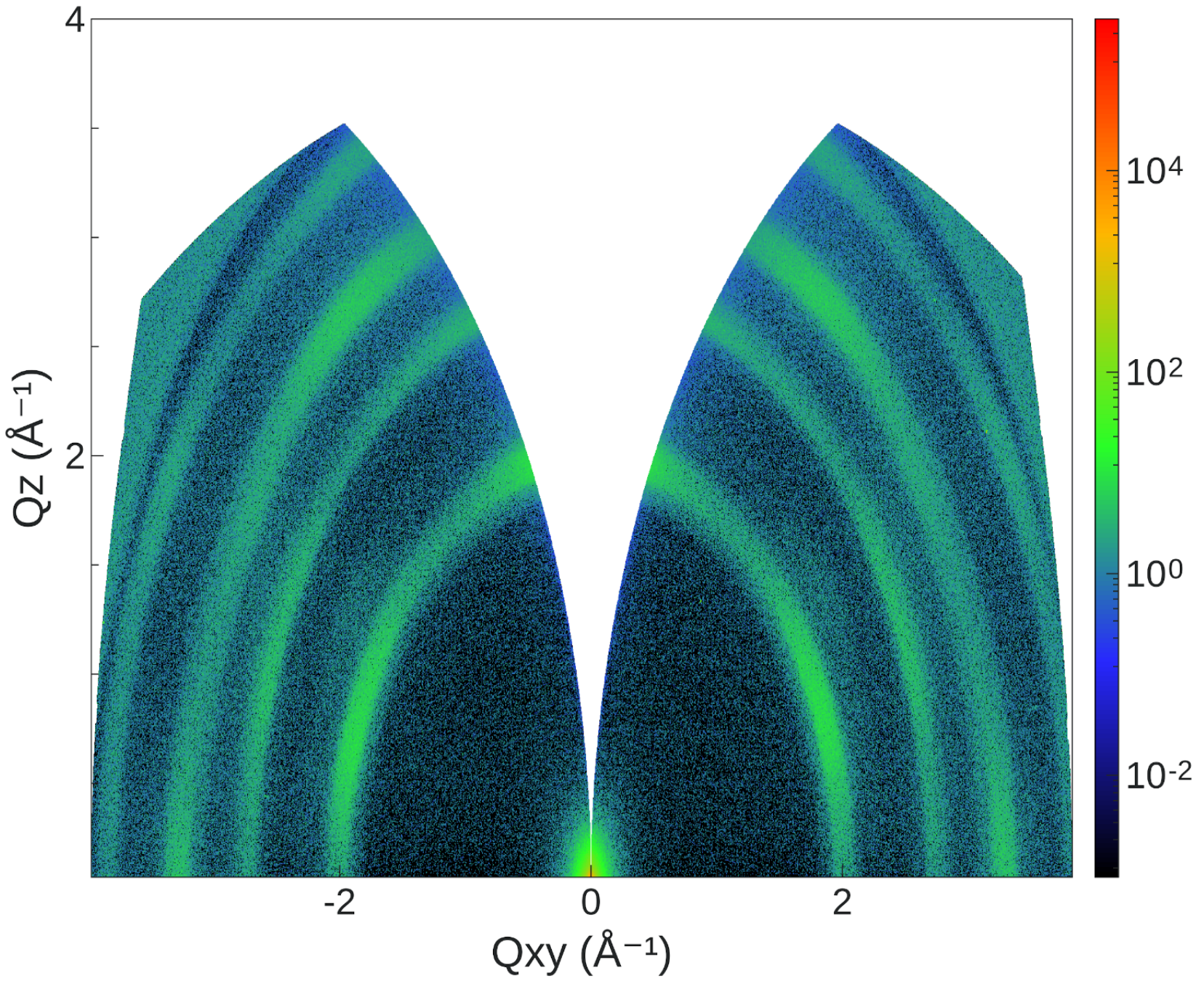

**FIG. S5**. GIWAXS pattern of 200 nm of ZnS on polycrystalline sputtered Pt on silicon. Polycrystalline rings at Q = 1.9 $Å^{-1}$ and 3.31 $Å^{-1}$ correspond to ZnS texturing in 100 and 112 families of reflections, respectively. Polycrystalline rings at Q = 2.77 $Å^{-1}$ and 3.2 $Å^{-1}$ correspond to Pt texturing in 111 and 002 families of reflections, respectively.

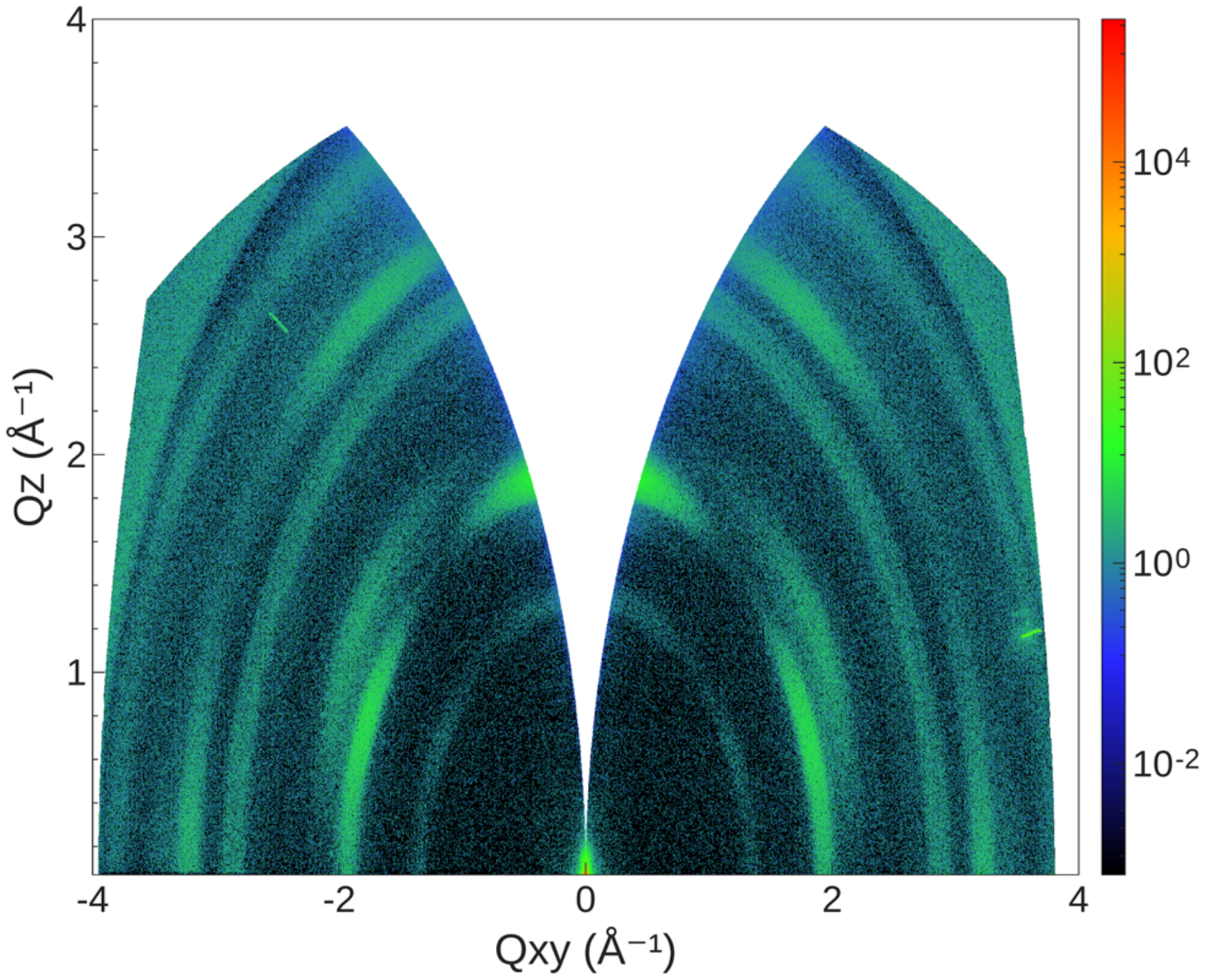

**FIG. S6.** GIWAXS pattern of 100 nm of $Y_2O_3$ on HF BOE etched silicon grown by PLD. The polycrystalline rings are indexed as follows: 112 family of reflections at Q = 1.45 Å$^{-1}$, 222 family of reflections at Q = 2.05 Å$^{-1}$, 224 family of reflections at Q = 2.90 Å$^{-1}$, 134 family of reflections at Q = 3.02 Å$^{-1}$, 125 family of reflections at Q = 3.25 Å$^{-1}$, and the 226 family of reflections at Q = 3.93 Å$^{-1}$.

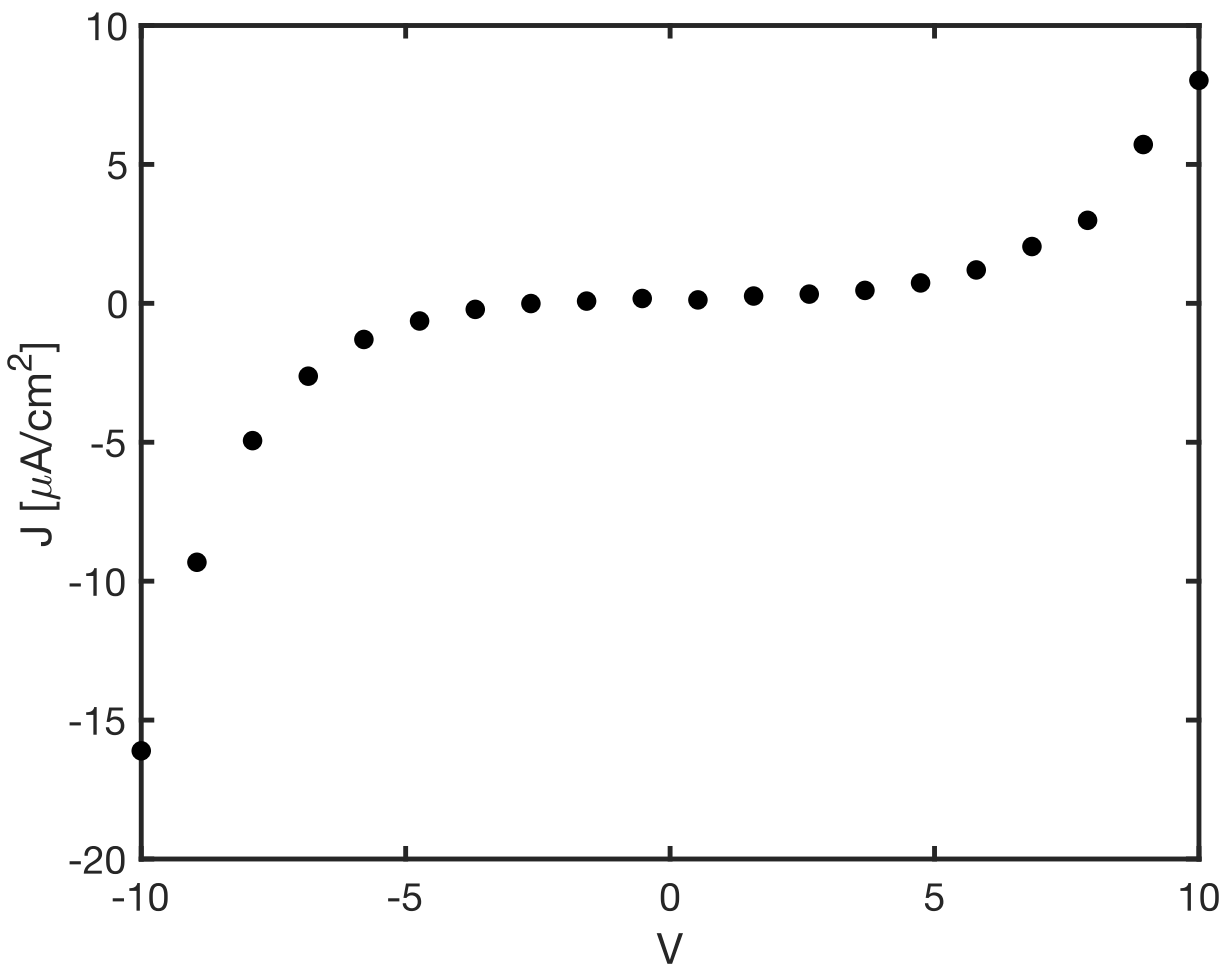

**FIG. S7.** Cross-plane J-V curve of 250 nm ZnS on Pt on heavily doped silicon.

## ACKNOWLEDGMENTS

This work was supported in part by the Taiwan Semiconductor Manufacturing Company (TSMC) Joint Development Program. C. W acknowledges the National Science Foundation of the United States under grant number DGE-1842487. The modification to the growth system to enable hybrid PLD was supported by an Air Force Office of Scientific Research grant no. FA9550-22-1-0117. The authors gratefully acknowledge the use of facilities at the Core Center for Excellence in Nano Imaging at University of Southern California for the results reported in this manuscript.

# AUTHOR DECLARATIONS

**Conflicts of Interest *(required)***

The authors declare no conflict of interest.

**Author Contributions *(if applicable)***

Statements about author contributions can be added here if needed.

A.T. Priyoti and G. Anilkumar performed electrical measurements. M. Surendran, K. Lin and A. Butler assisted with thin film growth. C. H. Wu, C. C. Wang, C. C. Kuo, R. Kapadia, H. Kumasubramanian and X. Bao reviewed the manuscript. J. Ravichandran provided input on research direction and reviewed the manuscript.

# DATA AVAILABILITY

The data that support the findings of this study are available from the corresponding author upon reasonable request.